\documentclass[vecphys]{svmult}

\usepackage{makeidx}         
\usepackage{graphicx}        
\usepackage{multicol}        
\usepackage{cite}            
\usepackage[bottom]{footmisc}
\usepackage{amsfonts,amsmath}

\makeindex             

\newcommand*{\be}{\begin{equation}}
\newcommand*{\ee}{\end{equation}}
\newcommand*{\bto}{\mbox{$\stackrel{\mathcal{B}}{\longrightarrow}$}}
\newcommand*{\cc}[1]{\bar{#1}}
\newcommand*{\re}[1]{\operatorname{Re}{#1}}
\newcommand*{\im}[1]{\operatorname{Im}{#1}}

\begin{document}

\title{Many-fermion wave functions: structure and examples}
\titlerunning{Many-fermion wave functions}
\author{D. K. Sunko}
\institute{Department of Physics, Faculty of Science, University of Zagreb
\texttt{dks@phy.hr}
}

\maketitle

\begin{abstract}
Many-fermion Hilbert space has the algebraic structure of a free module generated by a finite number of antisymmetric functions called \emph{shapes}. Physically, each shape is a many-body vacuum, whose excitations are described by symmetric functions (bosons). The infinity of bosonic excitations accounts for the infinity of Hilbert space, while all shapes can be generated algorithmically in closed form. The shapes are geometric objects in wave-function space, such that any given many-body vacuum is their intersection. Correlation effects in laboratory space are geometric constraints in wave-function space. Algebraic geometry is the natural mathematical framework for the particle picture of quantum mechanics. Simple examples of this scheme are given, and the current state of the art in generating shapes is described from the viewpoint of treating very large function spaces.
\end{abstract}


\section{Quantum mechanics and algebraic geometry}
\label{intro}

The standard textbook picture of quantum mechanics is that one-body wave functions represent possible states of individual particles, while many-body wave functions are constructed from the one-body functions by respecting indistinguishability for a \emph{given} number of particles, leading, in the case of fermions, to the well-known Slater determinants. This picture is called the \emph{particle picture} of quantum mechanics. In advanced textbooks a \emph{field picture} is introduced, which corresponds, as Dirac put it, to a ``deeper reality,'' meaning it refers from the outset to an infinite number of degrees of freedom.

The particle picture has a deeper reality of its own. Any wave function of $N$ identical fermions in $d$ dimensions may be written~\cite{Sunko16-1}
\be
\Psi=\sum_{i=1}^D\Phi_i\Psi_i,
\label{scheme}
\ee
where the $\Psi_i$ are antisymmetric and $\Phi_i$ are symmetric functions of $N$ particle coordinates. If the $\Phi_i$ were $c$-numbers, the $\Psi$ in Eq.~\eqref{scheme} would form a $D$-dimensional vector space. As it stands, it presents a finitely generated free module, where $D=N!^{d-1}$ is the dimension of the module (number of its generators). The $\Psi$ still belong to the full infinite-dimensional Hilbert space spanned by all Slater determinants, because of the additional degrees of freedom in the symmetric functions $\Phi_i$.

The scheme~\eqref{scheme} has an important geometric interpretation, which was discovered by Menaechmos in his construction of the cube root $\sqrt[3]{a}$. He interpreted the original observation by Hippocrates of Chios, namely
\be
y=x^2\quad\&\quad y^2=ax \implies x^4=ax,
\ee
to mean that the solution of $x^3-a=0$ could be found by intersecting two parabolas. This insight is fundamental to algebraic geometry: to represent an unknown object as an intersection of known objects. Extending this idea, Omar Khayyam found 19 classes of cubics by constructing various intersections of conics to solve them. From a modern viewpoint, due principally to Hilbert, his classes may be presented as \emph{ideals} generated by second-degree polynomials in $x$ and $y$, e.g.
\be
R=P\cdot(x^2-y)+Q\cdot(ax-y^2),
\label{PQ}
\ee
where $P$ and $Q$ are arbitrary polynomials in $x$ and $y$. This equation has the same structure as Eq.~\eqref{scheme}. Its defining characteristic is that simultaneous zeros of the generators are necessarily zeros of all members $R$ of the ideal.

Shapes are to a fermion many-body wave function what conics are to cubics. They are generators of all solutions to the $N$-fermion wave equation which respect the Pauli principle. Like the conics, they are not arbitrary functions, and also like the conics, there is a finite number of them. The efficient generation of shapes is the subject of current research efforts, described in the second part of this chapter.

\subsection{Technicalities}
\label{tech}
In order to implement the scheme~\eqref{scheme} most simply, a technical step is necessary. 
The Bargmann transform~\cite{Bargmann61} reads
\be
\mathcal{B}[f](t)=\frac{1}{\pi^{1/4}}
\int_{\mathbb{R}}dx\,
e^{-\frac{1}{2}\left(t^2+x^2\right)+xt\sqrt{2}}f(x)
\equiv F(t).
\ee
Here $f\in\mathrm{L}^2(\mathbb{R})$ and $F\in\mathrm{F}(\mathbb{C})$, the \emph{Bargmann space} of entire functions $F:\mathbb{C}\to\mathbb{C}$ such that
\be
\int_{\mathbb{C}}|F(t)|^2d\lambda(t)<\infty,
\ee
where
\be
d\lambda(t)=\frac{1}{\pi}e^{-|t|^2}d\re{t}\,d\im{t},\quad
\int_{\mathbb{C}}d\lambda(t)=1.
\ee
The inverse Bargmann transform is then
\be
\mathcal{B}^{-1}[F](x)=\frac{1}{\pi^{1/4}}
\int_{\mathbb{C}}d\lambda(t)\,
e^{-\frac{1}{2}\left(\cc{t^2}+x^2\right)+x\cc{t}\sqrt{2}}F(t),
\ee
where the bar denotes complex conjugation. 

Specifically, the Bargmann transform of Hermite functions $\psi_n(x)$ is
\be
\mathcal{B}[\psi_n](t)=\frac{t^n}{\sqrt{n!}}.
\label{barg}
\ee
It has the algebraically important property that quantum numbers (state labels) $n$ add when single-particle wave functions are multiplied, $t^nt^m=t^{n+m}$. Because $n!m!\neq (n+m)!$, one must use unnormalized single-particle wave functions $t^n$, with scalar product
\be
(t^n,t^m)=\int_{\mathbb{C}}\cc{t}^nt^md\lambda(t)=n!\,\delta_{nm}.
\label{scalp}
\ee
In three dimensions, the Hermite functions in $x$, $y$, and $z$ are mapped onto Bargmann-space variables $t$, $u$, and $v$, respectively. For $N$ particles, the variables acquire indices, e.g. $t_i$, with $i=1,\ldots,N$.

The technical advantage of Bargmann space is that the factorizations $\Phi_i\Psi_i$ in Eq.~\eqref{scheme} can be interpreted literally, as factorizations of polynomials. The same would not be so easy in laboratory space, where quantum numbers are indices of special functions, which have opaque properties under multiplication. One should bear in mind, however, that the free-module structure~\eqref{scheme} is an intrinsic feature of Hilbert space, irrespective of representation.

Another important technicality is that the generating function (Hilbert series~\cite{Sturmfels08}) which counts the shapes is known~\cite{Sunko16-1}. For $N$ fermions in $d$ dimensions, it is a polynomial $P_d(N,q)$, which satisfies \emph{Svrtan's recursion}
\begin{equation}
NP_d(N,q)=\sum_{k=1}^{N}(-1)^{k+1}\left[C^{N}_{k}(q)\right]^dP_d(N-k,q),
\label{svrtrec}
\end{equation}
where
\be
C^N_k(q)=\frac{(1-q^N)\cdots (1-q^{N-k+1})}{(1-q^k)}
\ee
is a polynomial, and $P_d(0,q)=P_d(1,q)=1$. For example, $P_2(3,q)=q^2+4q^3+q^4$, meaning that, of the $D=3!^{2-1}=6$ shapes of $N=3$ fermions in $d=2$ dimensions, one is a second-degree polynomial, four are third-degree, and one is fourth-degree.

\section{Simple examples~\cite{Rozman20}}
\label{exam}

\subsection{The fractional quantum Hall effect}
\label{fqhe}

A particular example of the free-module structure~\eqref{scheme} has been observed in the context of the $d=2$ fractional quantum Hall effect (FQHE), albeit without noticing its generality. Adopting the notation of Ref.~\cite{Laughlin90} for this subsection, one of the six shapes counted by $P_2(3,q)$ above is the (second-degree) ground-state Slater determinant,\footnote{The localization terms $\exp(-x_1^2/2-\ldots)$ are dropped for clarity.}
\be
\left|
\begin{matrix}
x_1 & x_2 & x_3\\
y_1 & y_2 & y_3\\
1 & 1 & 1
\end{matrix}
\right|,
\label{gs}
\ee
which is clearly not analytic in terms of the variables $z_j=x_j-iy_j$. One combination of the four third-degree shapes found in Ref.~\cite{Sunko16-1} is
\be
\Psi_0=
\left|
\begin{matrix}
z_1^2 & z_2^2 & z_3^2\\
z_1 & z_2 & z_3\\
1 & 1 & 1
\end{matrix}
\right|=(z_1-z_2)(z_1-z_3)(z_2-z_3),
\ee
while the other three involve terms with $\bar{z}_j$ on one or both rows of the determinant. The sixth (fourth-degree) shape goes into itself under the exchange $x_j\leftrightarrow y_j$, like the ground state~\eqref{gs}, so it is not analytic in the $z_j$ either.

On the other hand, Laughlin's $N=3$ wave function for the FQHE, Eq.~(7.2.12) of Ref.~\cite{Laughlin90}, contains a factor
\be
(z_a+iz_b)^{3m}-(z_a-iz_b)^{3m}=\Phi_m(z_1-z_2)(z_1-z_3)(z_2-z_3)=\Phi_m\Psi_0,
\label{factor}
\ee
where
\be
z_a=\frac{1}{\sqrt{6}}(z_1+z_2-2z_3),\quad
z_b=\frac{1}{\sqrt{2}}(z_1-z_2),
\ee
and $\Phi_m$ is a symmetric polynomial in the $z_j$. The factorization~\eqref{factor} is well known and easy to prove directly, which brings Laughlin's wave function into the scheme~\eqref{scheme}. This correspondence proves, by enumeration, Laughlin's conjecture~\cite{Laughlin90} that there is only one vacuum for $N=3$ and $d=2$ which satisfies the analyticity constraint.

\subsection{Two electrons in a quantum dot}
\label{qdot}

Two identical fermions in a three-dimensional harmonic potential are the simplest model of a finite system. It is easy to show that the Bargmann-space angular momentum operator has the same form as in laboratory space,
\be
L_z=-i(x\partial_y-y\partial_x) \bto -i(t\partial_u-u\partial_t)
\equiv \mathfrak{L}_v,
\ee
and cyclically. Therefore, solid harmonics in Bargmann space are the same polynomials as in real space, with $(x,y,z)$ simply replaced by $(t,u,v)$.

The generating function for this case is $P_3(2,q)=3q+q^3$. The ground-state triplet is a vector in Bargmann space just as in laboratory space,
\be
\vec{\Psi}=(t_1-t_2,u_1-u_2,v_1-v_2)=(\Psi_1,\Psi_2,\Psi_3),
\label{psivec}
\ee
while the fourth shape, appearing in the second-excited shell, is
\be
\widetilde{\Psi}=\Psi_1\Psi_2\Psi_3,
\label{psitil}
\ee
geometrically a pseudoscalar (signed volume). Introducing a vector of symmetric functions along the same lines,
\be
\vec{e}=(t_1+t_2,u_1+u_2,v_1+v_2)=(\eta_1,\eta_2,\eta_3)
\ee
the first-excited shell is spanned by $9$ vectors $\eta_i\Psi_j$, which is the scheme~\eqref{scheme} again. Knowing the form of the solid harmonics, it is easy to recast the scheme in rotationally invariant combinations. E.g., $\vec{e}\cdot\vec{\Psi}$ is a scalar, while $(-\eta_1+i\eta_2)(-\Psi_1+i\Psi_2)=e_{11}\Psi_{11}$ is a state with total angular momentum and projection $l=m=2$.

The second-excited shell is more interesting. In addition to squares $\eta_i^2$, there appear a new type of symmetric-function excitations, the discriminants
\be
\Delta_i=\Psi_i^2,\quad i=1,2,3,
\ee
which are excitations of \emph{relative} motion. There is a total of $10$ excitations involving relative motion alone, corresponding to the one-body oscillator states with three quanta. They are spanned by the fourth shape $\widetilde{\Psi}$ in addition to the nine states $\Delta_i\Psi_j$. Of the latter, one can easily construct a vector triplet,
\be
(\Delta_1+\Delta_2+\Delta_3)\vec{\Psi},
\label{vectrip}
\ee
noting that the sum of discriminants is a rotational scalar like $r^2$. The remaining $7$ states constitute a rotational septiplet $\Psi_{3m}$ with $l=3$, where $\Psi_{33}=\Psi_{11}^3$ and $\widetilde{\Psi}$ is embedded in the $m=\pm 2$ states:
\be
\widetilde{\Psi}\sim \Psi_{32}-\Psi_{3,-2}.
\ee
Even in this simplest possible example of a finite system, there appear two bands in the spectrum, because there are two shapes constraining the motion: the vector $\vec{\Psi}$, and the pseudoscalar $\widetilde{\Psi}$. All states in the spectrum can be classified according to whether they contain $\widetilde{\Psi}$ or not. The classification of finite-system spectra into bands is very like Omar Khayyam's classification of cubics by conics: bands are ideals generated by the shapes.

This example is quite revealing of the \emph{kinematic} (``off-shell'') nature of the constraint~\eqref{scheme}. The classification into bands is traditionally presented in the context of dynamics, i.e.\ some concrete equations of motion. Here one sees that bands are qualitative manifestations of geometric constraints in wave-function space, essentially many-body effects of the Pauli principle.

\section{Large spaces}
\label{large}

Simulations of strongly correlated systems must contend with the well-known fermion sign problem~\cite{Hirsch85}: it is not known in general how to update a many-fermion wave function consistently with an initial phase convention. Variational approaches avoid this problem, but at the price of limiting the possible range of wave functions to the form of the initial \emph{ansatz}.

The shape approach has the potential to obviate both problems. Because the number of shapes is finite, and they can be generated algorithmically in closed form, the expression~\eqref{scheme} is effectively a variational expression which is guaranteed to exhaust the whole wave-function space. It is natural to recast this program in probabilistic language, because the spaces involved are very large, so it is generally impossible to have a complete expression like~\eqref{scheme} stored in memory. The principal line of research into shapes at present is to generate an arbitrary shape with equal \emph{a priori} probability. In the remainder of this chapter, the current state of these efforts will be briefly presented. The principal results have not been published anywhere so far. In particular, the algorithm described in Ref.~\cite{Sunko16-3} is superseded here. From now on, the presentation is limited to the case of three dimensions.

\subsection{The size of the space}

The very large number of shapes $N!^2\sim N^{2N}$ must be put in perspective. The highest shape is unique and has degree $G=3N(N-1)/2$~\cite{Sunko16-3}. The total number of one-particle states up to the $G$-th oscillator shell is the sum of the shell degeneracies up to it:
\be
\sum_{n=0}^G\binom{n+2}{2}=\binom{G+3}{3}\sim N^6,
\ee
so for $N$ particles in the first $G$ shells the total number of states is
\be
\binom{\sim N^6}{N}\sim N^{6N}.
\ee
Even though the number of shapes is unimaginably large, it is vanishingly small in comparison with the total number of states spanning the same range of oscillator shells. Beyond the $G$-th shell, no new shapes appear, so the fraction of shapes in the total space can be made small at will.

These considerations open the way to structured simulations of large spaces, in which one optimizes in the shape subspace before taking other states into account. It amounts to using Eq.~\eqref{scheme} with $\Phi_i\in\mathbb{C}$ as a reduced variational expression in a first step. Such an approach makes physical sense because the nodal surface of the ground-state many-body function is expected to be an intersection of shapes alone: if the $\Phi_i$ introduced new nodes, these would correspond to excited states. [In Omar Khayyam's scheme~\eqref{PQ}, $R$ can similarly have roots which are not solutions of the cubic, because of $P$ and $Q$.] Thus one expects that corrections due to $\Phi_i\notin\mathbb{C}$ in the second step will change the geometry but not the topology of the ground-state nodal surface.

\subsection{The structure of shapes in three dimensions}

The highest-degree shape $S_N$ of $N$ identical fermions is a product of three 1D ground-state Slater determinants~\cite{Sunko16-3} $\tilde{\Delta}_N(t)$, cf. Eq.~\eqref{psitil},
\be
S_N=\tilde{\Delta}_N(t)\tilde{\Delta}_N(u)\tilde{\Delta}_N(v).
\label{SN}
\ee
Slater determinants in Bargmann space are Vandermonde forms
\be
\tilde{\Delta}_N(t)=
\left|\begin{matrix}
t_1^{N-1}/(N-1)!&\cdots&t_N^{N-1}/(N-1)!\\
\vdots&&\vdots\\
t_1/1!&\cdots&t_N/1!\\
1/0!&\cdots&1/0!
\end{matrix}\right|=
\prod_{1\le i<j\le N} \frac{t_i-t_j}{j-i}.
\ee
The normalization with factorials is for later convenience. Now it can be proven that symmetrized derivatives of $S_N$ are shapes. Denoting a derivative with respect to a variable with a capital letter, e.g.\ $T_i$ for $d/dt_i$, a symmetrized derivative is denoted by parentheses:
\be
(T^aU^bV^c)=\sum_{i=1}^N T_i^aU_i^bV_i^c,
\ee
which is a \emph{word} with $a+b+c$ commuting letters $T,U,V$, where $a,b,c\ge 0$ are integers. A product of words is a \emph{sentence}:
\be
(T^aU^bV^c)\cdots(T^xU^yV^z).
\label{sentence}
\ee
A sentence is a symmetric polynomial in the derivatives, call it $P(T,U,V)$. The defining characteristic of a shape is that it is orthogonal to states of the form $\Phi\Psi$, where $\Psi$ is a shape, and $\Phi\neq 1$ a symmetric polynomial corresponding to a bosonic excitation of $\Psi$ --- these were called ``trivial states'' in Ref.~\cite{Sunko16-1}.

The claim is that $P(T,U,V)S_N$ is a shape if $S_N$ is a shape. The proof is standard~\cite{Bergeron12}:
\be
(P(T,U,V)S_N,\Phi\Psi)=(S_N,\Phi P(t,u,v)\Psi)=0,
\ee
where $P(t,u,v)$ is the symmetric polynomial obtained by replacing all derivatives in $P(T,U,V)$ with the corresponding variables. The first equality follows because variables and derivatives are conjugate in the scalar product~\eqref{scalp}, and the second because the product $P(t,u,v)\Psi$ is multiplied by $\Phi\neq 1$, which gives a trivial state, no matter how $P(t,u,v)\Psi$ itself is resolved in the scheme~\eqref{scheme}.

\begin{figure}[t]
\centering
\includegraphics*[width=.7\textwidth]{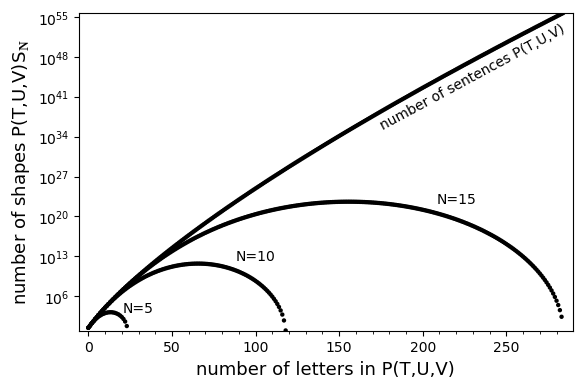}
\caption[]{The number of shapes approaches the number of sentences asymptotically when the number of particles $N\to\infty$, as shown here for $N=5,10,15$}
\label{asymptotics}       
\end{figure}

For example, $(TU)S_2=2(v_1-v_2)$, cf. Eq.~\eqref{psivec}. This approach is far more efficient than the originally published one~\cite{Sunko16-1}, because it generates shapes directly, without the need to generate the much larger oscillator-shell spaces in which they are embedded. The main open problem at present is that the number of sentences is still much larger than the number of shapes. This redundancy is illustrated in Fig.~\ref{asymptotics}. The number of sentences\footnote{With all words having at least two distinct letters, because $(T^a)\Delta_N(t)=0$~\cite{Sunko16-3}.} is equal to the number of shapes only when the total number of letters in a sentence is equal or smaller than $N$, the number of particles. Higher and higher derivatives of $S_N$ eventually reach the ground state for any finite $N$, after which the number of shapes is zero, as observed in the figure.

\subsection{Shape notation for a given number of particles}

The sentence notation is adapted to an arbitrary number of particles. Any sentence acting on $S_N$ will give a shape (or zero), irrespective of the value of $N$. A notation focused on a given finite $N$ is introduced for the redundancy problem now.

Because derivatives act on columns of $\tilde{\Delta}_N$, the factorials allow a compact expression for the outcome of such operations:
\be
[n_1\ldots n_N]=\left|\begin{matrix}
t_1^{n_1}/n_1!&\cdots&t_N^{n_N}/n_N!\\
t_1^{n_1-1}/(n_1-1)!&\cdots&t_N^{n_N-1}/(n_N-1)!\\
\vdots&&\vdots\\
0 \mbox{ or } 1 &\cdots& 0 \mbox{ or } 1
\end{matrix}\right|.
\ee
For example, $\tilde{\Delta}_3(t)=[2\,2\,2]$ and $d\tilde{\Delta}_3(t)/d t_1=[1\,2\,2]$. This determinant is a slight generalization of the so-called \emph{confluent Vandermonde form}. Writing the three terms in the product~\eqref{SN} vertically in the order $t,u,v$, the variables can be left implicit, e.g.
\begin{multline}
\begin{matrix}
[0\,1\,2] \\
[1\,1\,2] \\
[1\,2\,2]
\end{matrix}=
\left|\begin{matrix}
1&t_2&t_3^2/2\\
0&1&t_3\\
0&0&1
\end{matrix}\right|\cdot 
\left|\begin{matrix}
u_1&u_2&u_3^2/2\\
1  &1  &u_3\\
0  &0  &1
\end{matrix}\right|\cdot
\left|\begin{matrix}
v_1&v_2^2/2&v_3^2/2\\
1  &v_2  &v_3\\
0  &1  &1
\end{matrix}\right|=\\
1\cdot (u_1-u_2)\cdot (v_2-v_3)(v_1-v_2/2-v_3/2).
\end{multline}
A symmetrized derivative is obtained by summing all possible arrangements of columns, e.g.
\be
\left\{\begin{matrix}
1\,1\,2 \\
1\,1\,2 \\
2\,2\,2
\end{matrix}
\right\}
=
\begin{matrix}
[1\,1\,2] \\
[1\,1\,2] \\
[2\,2\,2]
\end{matrix}
+\begin{matrix}
[1\,2\,1] \\
[1\,2\,1] \\
[2\,2\,2]
\end{matrix}
+\begin{matrix}
[2\,1\,1] \\
[2\,1\,1] \\
[2\,2\,2]
\end{matrix}.
\label{prim2simb}
\ee
The curly brackets denote a \emph{shape symbol}, or symbol for short. For a given $N$, sentences and symbols are related linearly, e.g.
\be
(TU)^2S_3=
\left\{\begin{matrix}
0\,2\,2 \\
0\,2\,2 \\
2\,2\,2
\end{matrix}\right\}
+
2\left\{\begin{matrix}
1\,1\,2 \\
1\,1\,2 \\
2\,2\,2
\end{matrix}\right\},
\ee
where the first term corresponds to the $i=j$ terms in the double sum $(TU)^2$, i.e.\ the word $(T^2U^2)$, while the second comes from the parts with $i<j$ and $i>j$. The expansion of each sentence into symbols generates at least one distinct symbol (the second one in the example above), so the symbols are just as good a basis for the shapes as the sentences.

Constraints can be deduced for the symbols from the underlying determinants. For example, a symbol will be zero if all entries on any row are less than $N-1$, because the corresponding determinant then has a row of zeros. These constraints reduce the number of symbols with respect to the number of sentences, which is practical enough for smaller problems, but there still remain many more symbols than shapes of a given degree in general.

If one could find a set of constraints which were both \emph{efficient} in the sense that they allowed every distinct shape to be generated exactly once, and \emph{operative} in the sense that they could be implemented in polynomial time in $N$, the problem of generating all shapes with equal \emph{a priori} probability would be solved. Such progress is unlikely to happen by trial and error, because the redundancy problem has a structure and physical meaning of its own, described in the next section.

\subsection{Syzygies and the fermion sign problem}

Sentences can be classified according to the total powers of $T$, $U$, and $V$ appearing in them:
\be
P(T,U,V)\simeq (T^aU^bV^c).
\ee
As an example, take $N=4$ particles and generate all shapes $P(T,U,V)S_4$ with $(a,b,c)=(2,2,4)$. There are exactly five of them, so the whole subspace in the class of $(T^2U^2V^4)$ can be spanned by five distinct symbols, such as
\be
{\scriptstyle
\left\{\begin{matrix}
2\,2\,3\,3 \\
2\,2\,3\,3 \\
1\,1\,3\,3
\end{matrix}\right\},\,
\left\{\begin{matrix}
2\,2\,3\,3 \\
1\,3\,3\,3 \\
1\,3\,2\,2
\end{matrix}\right\},\,
\left\{\begin{matrix}
2\,2\,3\,3 \\
1\,3\,3\,3 \\
1\,1\,3\,3
\end{matrix}\right\},\,
\left\{\begin{matrix}
1\,3\,3\,3 \\
3\,1\,3\,3 \\
1\,3\,2\,2
\end{matrix}\right\},\,
\left\{\begin{matrix}
1\,3\,3\,3 \\
2\,2\,3\,3 \\
1\,3\,2\,2
\end{matrix}\right\}.
}
\label{basis}
\ee
Practically, they are found as follows. The highest shape $S_4$ has $[3\,3\,3\,3]$ on each row, so the row-sum is $12$. A derivative such as $(T^2U^2V^4)$ subtracts $2$ from the first and second rows, and $4$ from the third, so all shapes in the same class can be generated by symbols whose row-sums are $10$, $10$, and $8$ respectively. When all so-far known constraints and symmetries are implemented, $21$ symbols are allowed. One expands them one by one in the underlying variables $t_i,u_i,v_i$ until five distinct ones are found. The dimension of the subspace (five) is known in advance from a refinement of Svrtan's recursion: in the formula~\eqref{svrtrec}, drop the $(-1)^{k+1}$ and replace (for $d=3$)
\be
\left[C^{N}_{k}(q)\right]^3\rightarrow
C^{N}_{k}(T)\cdot C^{N}_{k}(U)\cdot C^{N}_{k}(V).
\ee
With these modifications, the recursion gives a polynomial $B(N,T,U,V)$, such that the coefficient of $T^aU^bV^c$ is the dimension of the shape subspace of class $(T^aU^bV^c)$. In this particular example,
\be
B(4,T,U,V)=\ldots + 5\cdot T^2U^2V^4 + \ldots,
\ee
so the algorithm can stop as soon as five distinct symbols are found. Among all the $21$ symbols, one finds relations such as
\be
\left\{\begin{matrix}
2\,2\,3\,3 \\
1\,3\,3\,3 \\
3\,1\,2\,2
\end{matrix}\right\}+
\left\{\begin{matrix}
2\,2\,3\,3 \\
1\,3\,3\,3 \\
1\,3\,2\,2
\end{matrix}\right\}-
\left\{\begin{matrix}
2\,2\,3\,3 \\
1\,3\,3\,3 \\
1\,1\,3\,3
\end{matrix}\right\}=0.
\label{syz}
\ee
In the theory of invariants, such relations are called \emph{syzygies}~\cite{Sturmfels08}. They appear characteristically as polynomial expressions in some determinants, which vanish when the determinants are expanded in the underlying variables. When the determinants express geometrical constraints, a syzygy shows which constraints imply one another. An example is Desargues' theorem: adjusting some lines to intersect is the same as adjusting some points to be collinear. The syzygy~\eqref{syz} is a challenge to begin developing such geometric insights for constraints induced by the Pauli principle in wave-function space. Notice, for example, that this particular syzygy is one-dimensional, because the first two rows in all three symbols are the same.

Syzygies are algebraic expressions of the fermion sign problem. It appears in Eq.~\eqref{syz} as three different wave functions which interfere destructively because there are actually only two distinct functions. Generating only distinct symbols is solving the sign problem, because it fixes the expression~\eqref{scheme} as a variational \emph{ansatz}.

Syzygies of shape symbols have physical meaning. Sentences represent deexcitations (loss of quanta) of the high-lying special state $S_N$. In a fast deexcitation cascade starting with that state, expressions like~\eqref{syz} are physical interference effects between different branches of the cascade: not to include them would get the branching ratios wrong. Syzygies become a problem only when one wants to model the \emph{equilibrium} state, requiring that all distinct wave functions be included with equal \emph{a priori} probabilities.

Algebraically, syzygies generate their own ideal, the \emph{syzygy ideal} of the invariant ring. Calculations in the quotient ring \emph{modulo} the syzygy ideal are real in the sense that one is not inadvertently manipulating zeros. Methods developed for such computations are of two kinds. One requires ``unpacking'' the determinantal expressions in terms of the underlying variables, such as the algorithm described above, which discovered the basis~\eqref{basis} and the syzygy~\eqref{syz}. The other is symbolic, in the sense that some rules at the level of the symbols themselves determine which expressions are allowed. This second type of method is the goal of the present research, because expanding the symbols in terms of the underlying variables rapidly becomes prohibitive when the number of fermions increases. It amounts to finding efficient constraints on the symbols, as discusssed in the previous section.

\section{Connection to field theory}

It has been pointed out before~\cite{Sunko16-1} that the excitations $\Phi_i$ in Eq.~\eqref{scheme} are counted in a manner strictly analogous to the quantization of the electromagnetic field. Namely, the excitations in the three directions in space are mutually independent. Every $\Phi_i$ can be expanded in monomials of the form $\Phi^x_j\Phi^y_k\Phi^z_l$, where each $\Phi^{x,y,z}$ is a symmetric one-dimensional function \emph{by itself}. A 3D excitation is just an iteration of 1D excitations.

The connection of these excitations to field theory is straightforward: just let the number of particles $N\to\infty$. This limit is implicit in the theory of symmetric functions, where one assumes by default that $N$ is fixed but may be arbitrarily large, as if the ``supply of variables'' were inexhaustible~\cite{Stanley99}. Physically, the heat capacity of the electromagnetic field increases without bound with temperature because the field has infinitely many degrees of freedom in a finite volume, so the supply of degrees of freedom (wave-numbers) is inexhaustible, even if not all are excited.

The important difference between the excitations $\Phi_i$ and field degrees of freedom is that the former have no zero-point energy, which resides in the shapes~\cite{Sunko16-1}. In the limit $N\to\infty$, the mapping of sentences to shapes (Fig.~\ref{asymptotics}) becomes one-to-one. Notably, all words in a sentence must have at least two distinct letters, because one-dimensional words $(T^a)$ give zero when acting on $S_N$~\cite{Sunko16-3}. While $S_N$ is a product of 1D forms, one needs to ``tie together'' the spatial directions in a word in order to obtain a shape in general. This property is in sharp contrast with the excitations, which factorize into one-dimensional functions $\Phi^{x,y,z}$ \emph{only}. A field-theoretic description of fermionic vacua should map to the space of (infinitely long) sentences~\eqref{sentence}, not to the excitations $\Phi_i$. It remains to be seen whether such a connection is possible.

\section{Discussion}

The success of the Slater determinant basis rests on the physical fact that the ground-state Slater determinant is a good starting point for the description of the ground state and low-lying excitations of many real systems. The simple algebraic structure of this determinant then provides the formal advantage, that one can manipulate a very large expression --- a Slater determinant of $N=10^{23}$ particles --- without ever having to ``unpack'' it in terms of the one-particle labels. This advantage is realized through the well-known formalism of second quantization.

The advantage disappears in strongly correlated cases, which are formally characterized as those where the initial \emph{ansatz} needs to be more complex than a single Slater determinant. These are known as configuration-interaction approaches in quantum chemistry, going back to the Heitler-London wave function. One can still use the second-quantized notation, but its cost-effectiveness is ultimately compromised by the \emph{physical} necessity to calculate in terms of the individual orbitals. The end point of such a deconstruction are quantum Monte Carlo (QMC) simulations, which rapidly become prohibitive with a larger number of particles.

The present work puts back some structure into the latter efforts, motivated by the desire to write a strongly correlated variational \emph{ansatz} without prejudice. A QMC calculation deals with a very large Hilbert space. However, not all states in that space are born equal. A minority, the shapes, are distinguished by the ability to act as vacuum states, or algebraically, as generators of the Hilbert space as a free module. These are natural generalizations of the ground-state Slater determinant. Learning to manipulate these states at the formal level, using only the shape symbols as labels, is the corresponding generalization of the second-quantized formalism. It promises to mobilize the apparatus of algebraic geometry and classical invariant theory for many-body problems with strong correlations.

The interpretation of the fermion sign problem in terms of syzygies is a case in point. The syzygies are a nuisance in equilibrium calculations, which is the sign problem, but they also have a real meaning. Physically, the sentences and symbols represent chains of de-excitation of the initial (highest) shape $S_N$. A syzygy like~\eqref{syz} means that the same wave function may be constructed by apparently different laboratory preparations, e.g.\ photon-loss sequences. In chemists' language, it indicates different synthetic pathways to the same entangled state.

The daunting size of the shape space, $N!^2$ in three dimensions, has two mitigating factors. One is that current simulations work in a much larger space by default, so one who is not afraid of $N^{6N}$ can hardly complain that $N^{2N}$ is too much. However, that leaves open the possibility that such heavy work for very large $N$ may be unnecessary, which leads to the second mitigating factor. Experience with actual strongly correlated systems --- the FQHE~\cite{Laughlin90}, large molecules~\cite{Nakatsuji15}, and modern functional materials~\cite{Hyowon14} --- indicates a hierarchichal organization of the wave function, in which comparatively few electrons create a correlated template, which is then extended across the system. With the developments described above, the shape formalism is already operational for $N\sim 5$, which is the number of identical fermions in a $d$ shell, so it is more applicable to real-world problems than appears from the present article.

Investigating the geometry of wave-function space seems worthy both as an end in itself and as a practical tool. It is a new open frontier of fundamental quantum mechanics.


\section{Acknowledgements}

I thank J.~Bon\v{c}a and S.~Kruchinin for the invitation to present these results at the NATO Advanced Research Workshop in Odessa. Conversations with M.~Primc and D.~Svrtan are gratefully acknowledged. This work was supported by the Croatian Science Foundation under Project No.~IP-2018-01-7828 and University of Zagreb Support Grant 20283207.

\raggedright
%


\printindex
\end{document}